\def\rfr#1{Eq. (\ref{#1})}

\def\derp#1#2{\rp{\partial{#1}}{\partial{#2}}}
\def\dert#1#2{\frac{{{d}}{#1}}{{{d}}{#2}}}              

\def\virg#1{``#1''}

\def\eqi{\begin{equation}}
\def\eqf{\end{equation}}
\def\eqia{\begin{eqnarray}}
\def\eqfa{\end{eqnarray}}
\def\rp#1#2{{#1\over#2}} \def\lb#1{\label{#1}}

\documentclass[reprint,showpacs,showkeys,amsmath,amssymb,aps]{revtex4-1}
\usepackage{graphicx}
\usepackage{dcolumn}
\usepackage{bm}
\usepackage{url}
\usepackage{amsmath,amsthm,amscd,amssymb}
\usepackage{latexsym,wasysym}
\usepackage{graphicx,epsfig}

\RequirePackage{color}

\begin{document}

\title{\textcolor{black}{Model-independent constraints on $r^{-3}$ extra-interactions from orbital motions}}

\author{L. Iorio}
\altaffiliation{Ministero dell'Istruzione, dell'Universit\`{a} e della Ricerca (M.I.U.R.)-Istruzione. Fellow of the Royal Astronomical Society (F.R.A.S.).
 International Institute for Theoretical Physics and
Advanced Mathematics Einstein-Galilei. Permanent address: Viale Unit$\grave{\rm a}$ di Italia 68
70125 Bari (BA), Italy.}
\email{lorenzo.iorio@libero.it}

\begin{abstract}
\textcolor{black}{Constraints on long-range  power-law modifications $U_{\rm pert}\propto r^{-3}$ of the usual  Newtonian gravitational potential $U_{\rm N}\propto r^{-1}$}   are inferred from orbital motions of well known artificial and natural bodies. \textcolor{black}{They can be interpreted in terms of a characteristic length $\ell$ which may be identified with, e.g., the anti-de Sitter (AdS) radius of curvature  ${\ell}$ in the  Randall-Sundrum (RS) braneworld model, although this not a mandatory choice. Our bounds, complementary to those from tabletop laboratory experiments,} do not rely upon more or less speculative and untested theoretical assumptions, contrary to other long-range RS tests proposed in astrophysical scenarios in which many of the phenomena adopted may depend on the system's composition, formation and dynamical history as well.
\textcolor{black}{Independently of the interpretation of $\ell$,} the perihelion precession of Mercury and its radiotechnical ranging from the Earth yield ${\ell} \lesssim 10-50\ {\rm km}$. Tighter bounds come from the perigee precession of the Moon, from which it can be inferred ${\ell} \lesssim 500-700\ {\rm m}$. The best constraints (${\ell} \lesssim 5\ {\rm m}$) come from the Satellite-to-Satellite Tracking (SST) range of the GRACE A/B spacecrafts orbiting the Earth: proposed follow-on of such a mission, implying a sub-nm s$^{-1}$ range-rate accuracy, may constrain ${\ell}$ at $\sim 10\ {\rm cm}$ level. Weaker constraints come from the double pulsar system (${\ell}\lesssim 80-100\ {\rm km}$) and from the main sequence star S2 orbiting the compact object in Sgr A$^{\ast}$ (${\ell}\lesssim 6.2-8.8\ {\rm AU}$).
\textcolor{black}{Such bounds on the length $\ell$, which must not necessarily be identified with the AdS radius of curvature of the RS model, naturally translate into constraints on an, e.g., universal coupling parameter $\mathcal{K}$ of the $r^{-3}$ interaction. GRACE yields $\mathcal{K}\leq 1\times 10^{16}\ {\rm m^5\ s^{-2}}$.}
\end{abstract}



\maketitle

\section{Introduction}
\textcolor{black}{In this paper, we deal with power-law modifications of the usual inverse-square law \cite{Fisch01,Adel03}.}

Extra dimensions, arising in string theory, supergravity, M-theory \cite{polchi} and string inspired higher dimensional theories such as
the brane-world models \cite{randall,rs,shiromizu}, have recently gained increasing importance in physics in the context of the search for a quantum-gravity theory. In particular, large, non-compactified extra dimensions could potentially
solve the long-lasting hierarchy problem \cite{arkadi,antoniadis,randall,rs}.
Indeed, if the standard model of particles and fields is restricted only on a
(3+1)-dimensional brane, whereas gravity is allowed to propagate
in the higher-dimensional bulk, the effective Planck scale
in the four-dimensional spacetime can be made significantly
larger than the electroweak scale, matching the
experimental requirements.

In the braneworld model by Randall and Sundrum (RS hereafter) \cite{rs},
our usual four-dimensional spacetime is a brane, which the standard model
fields are constrained to, embedded in a five-dimensional anti-de Sitter (AdS) spacetime. In it, the fifth spatial dimension can be infinite, with an AdS curvature scale ${\ell}$. Indeed, the RS model circumvents the need of compactifying all but the three observed spatial dimensions
by including a bound state of the massless graviton
on the brane \cite{rs} resulting from the curvature, rather than
the size, of the extra dimension. At distances $r\gg {\ell}$, the RS model
implies a correction $U_{\rm RS}$ to the Newtonian gravitational potential of a body of mass $M$ at second post-Newtonian order (2PN). It is given by
\eqi U_{\rm RS} = -k\rp{GM}{r}\left(\rp{{\ell}}{r}\right)^2,\lb{upert}\eqf
where $G$ is the Newtonian constant of gravitation, and $k$ can assume different values depending on the schemes of regularization adopted \cite{jung}. E.g., it can be  $k=1$ \cite{rs}, $k=1/2$ \cite{barvi}, and $k=2/3$ \cite{garriga}. The occurrence \textcolor{black}{of a correction to the Newtonian potential of the form of} \rfr{upert}, \textcolor{black}{which, however, is not necessarily limited to the RS model, being common to a wide class of power-law interactions \cite{Adel03}}, is important since, although it is only gravity
that feels the presence of the extra dimensions, \rfr{upert} allows for detectable effects on our (3+1)-brane
that can be used in constraining the properties of the bulk.
For upper limits on the brane parameter of other braneworld models, see, e.g., \cite{bobo1,bobo2}.

Several large-scale tests of the RS model \cite{rs}  have been proposed so far in astrophysical scenarios; they claimed bounds on ${\ell}$ at  $\sim 1-10\ \mu$m level, which is the same order of magnitude reached in laboratory-scale experiments \cite{adelberger,russi}. Anyway, such constraints  typically depend on the particular
interpretation of astrophysical observations, and
suffer from large systematic effects whose accurate knowledge is often lacking. Moreover,
they are often quite model-dependent in the sense that they heavily rely upon theoretical assumptions which are still speculative since they have not yet been tested independently with a variety of different phenomena, or have not yet been directly tested at all. In our opinion, it is true also for those tests \cite{psaltis,gnedin,johanssen,johanssen2} requiring the least amount of information like, e.g., the evaporation
of black holes \cite{traschen}. Indeed, they are based on the application of a concept like the anti–de Sitter space/conformal
field theory (AdS/CFT) duality \cite{adscft} in braneworld gravity
models that are asymptotically AdS (such as the RS models). Moreover, the consequent theoretical prediction for the black hole evaporation time \cite{evapo1,evapo2,evapo3}
depends only on assumptions regarding the braneworld
model and, in particular, on the validity and implementation
of the AdS/CFT correspondence in view of the \virg{no-hair} conjecture \cite{hair1,hair2}, which is itself speculative  and still awaits independent observational checks \cite{hair3,hair4}. Suffice it to say that the basis of
the previously cited calculation for the black hole evaporation time \cite{evapo1,evapo2,evapo3} has been recently challenged in Ref. \cite{fitz}.
In addition to such theoretical considerations at fundamental level, it must also be remarked that the astrophysical phenomena themselves used to constrain ${\ell}$, like the orbital evolution of  black-hole X-Ray binaries or the behavior of black holes in extragalctic clusters, are not lacking of uncertainties, both from a theoretical and observational point of view. E.g., they may crucially depend on the composition, formation and  dynamical history of the systems considered. Last but not least, black holes may well not exist at all \cite{ball,chapline}; e.g., even in the case of the compact object in Sgr A$^{\ast}$, there are not yet direct evidences that its $\sim 10^6 M_{\odot}$ mass is actually concentrated within its Schwarzschild radius $R_s = 0.084\ {\rm AU}$  \cite{balbi}. A signature for the absence of
event horizons was even looked for by the authors of Ref. \cite{barbieri}. In conclusion, to date, a definite proof for the existence of Kerr black holes is still lacking despite a wealth of observational evidence \cite{libro}.
Moreover, from a broader point of view, in correctly assessing the relevance of the laboratory-based tests of long-range modified models of gravity it should be mentioned that some authors \cite{envi1,envi2} pointed out the need of go to space as well since  such extra gravitational degrees of freedom   may have an environmental dependence.

Thus, in Sec. \ref{seconda} we will use well known and largely tested orbital motions (see Sec. \ref{perielio}) of some natural (Sec. \ref{pianeti}) and artificial (Sec. \ref{grazia})  bodies in the solar system  to infer constraints on
the AdS radius of
curvature which, if on the one hand, are not at $\mu$m level, on the other hand can certainly be considered  less speculative than those obtained in astrophysical contexts. Indeed, apart from the fact that there are no doubts about the existence and the properties of the planets of the solar system and of man-made Earth's satellites, the dynamical effects used to constrain ${\ell}$  are straightforwardly obtained from \rfr{upert}, without any additional hypothesis concerning untested phenomena. Moreover, competing effects acting as systematic errors are known with a comparatively much better accuracy. We notice that McWilliams \cite{gwaves} recently proposed to constrain ${\ell}$ via gravitational wave measurements in the solar system: LISA would allow to place bounds on the AdS radius of
curvature of the order of ${\ell}\sim 1\ \mu$m from the event rate of stellar black holes
inspiraling gravitationally into supermassive black holes, and of ${\ell}\lesssim  5\ \mu$m from the observation of
individual galactic binaries containing a stellar mass black
hole. In Sec. \ref{neutrona} we will also use  the well known, and extensively studied, double pulsar binary system and the main sequence  S2 star orbiting the compact object in Sgr A$^{\ast}$.
Finally, we stress that our results are not necessarily limited to the RS model \cite{rs}, being valid also for other theoretical schemes yielding  power-law interactions $\propto r^{-3}$ \textcolor{black}{\cite{moste,ferrer,ferrer2,dob,adelberger,adel09}}.
\section{RS long-term orbital effects and comparison with the observations}\lb{seconda}
\subsection{Analytical calculation of the secular precession of the pericenter}\lb{perielio}
The long-period effects caused by \rfr{upert} on the orbital motion of a test particle of mass $m$ can  be
computed perturbatively by adopting the Lagrange equations for the variation of the osculating Keplerian elements \cite{capderou}: their validity has been confirmed in a variety of independent phenomena.
Generally speaking, they imply the use of a perturbing function $\mathcal{R}$ which is the correction $U_{\rm pert}$ to the standard Newtonian monopole term.
In the case $U_{\rm pert}=U_{\rm RS}$, the average over one orbital revolution of the perturbing function $\mathcal{R}$ is straightforwardly obtained by using the true anomaly $f$ as fast variable of integration: it is
\eqi \left\langle\mathcal{R}\right\rangle = -\rp{GM k {\ell}^2}{ a^3\left(1-e^2\right)^{3/2}},\lb{pertfun}\eqf
where $a$ is the semimajor axis and $e$ is the eccentricity of the test particle's orbit.
From \rfr{pertfun} and  the Lagrange equation for variation of the longitude of pericenter $\varpi$ \cite{bertotti}
\eqi \left\langle\dert\varpi t\right\rangle = -\rp{1}{n_{\rm b} a^2}\left\{\left[\rp{\left(1-e^2\right)^{1/2}}{e}\right]\derp{\left\langle\mathcal{R}\right\rangle}e + \rp{\tan\left(I/2\right)}{\left(1-e^2\right)^{1/2}}\derp{\left\langle\mathcal{R}\right\rangle}I\right\}, \lb{lagra}\eqf
it turns out that $\varpi$ experiences a secular precession given by
\eqi \left\langle\dert\varpi t\right\rangle = \rp{3\left(GM\right)^{1/2} k{\ell}^2}{ a^{7/2} \left(1-e^2\right)^2}.\lb{precess}\eqf In \rfr{lagra}, $n_{\rm b}\doteq \sqrt{GM/a^3}$ is the Keplerian mean motion  and $I$ is the inclination of the orbital plane to the reference   $\{x,y\}$ plane. The longitude of pericenter $\varpi\doteq\Omega+\omega$ is a \virg{dogleg} angle since it is the sum of the longitude of the ascending node $\Omega$, which is an angle in the reference $\{x,y\}$ plane from a reference $x$ direction to the intersection of the orbital plane with the $\{x,y\}$  plane itself (the line of the nodes), and of the argument of pericenter $\omega$, which is an angle counted in the orbital plane from the line of the nodes to the point of closest approach, usually dubbed  pericenter. The precession of \rfr{precess}, which is an exact result in the sense that no a-priori assumptions on $e$ were made, agrees with the one obtained by Adkins and McDonnel in Ref. \cite{adkins} with a more cumbersome calculation. To facilitate a comparison between such two results, we note that, in general, the authors of Ref. \cite{adkins} work out the perihelion advance per orbit $\Delta\theta_p$: it corresponds to $\left\langle\dot\varpi\right\rangle P_{\rm b},$ where $P_{\rm b}=2\pi/n_{\rm b}$ is the orbital period. Moreover, in the potential energy $V(r)=\alpha_{-(j+1)}r^{-(j+1)}$ it must be posed $j=2$ and $\alpha_{-3}\rightarrow -GMmk{\ell}^2$, while in $\Delta\theta_p(-(j+1))$ of Eq. (38) in Ref. \cite{adkins} it must be set $L\rightarrow a(1-e^2),\ \chi_2(e)=6$. With such replacements, it can be shown that the advance per orbit of Eq. (38) in Ref. \cite{adkins} corresponds just to our precession in \rfr{precess}. Moreover, it turns out that the analytical result of \rfr{precess} is confirmed by a numerical integration of the equations of motion for Mercury with, say, ${\ell}=10^{-6}$ AU and $k=1/2$: both yield 5.4 milliarcseconds per century (mas cty$^{-1}$ hereafter). The choice of the numerical value adopted for ${\ell}$ is purely arbitrary, being motivated only by the need of dealing with relatively small numbers.
%
%
%
%
%
%
\subsection{Constraints from solar system planetary orbital motions}\lb{pianeti}
The corrections $\Delta\dot\varpi$ to the standard Newtonian-Einsteinian secular precessions of the longitudes of the perihelia are routinely used by independent teams of astronomers \cite{pitjeva,fienga} as a quantitative measure of the maximum size of any putative anomalous effect allowed by the currently adopted mathematical models of the standard solar system dynamics fitted to the available planetary observations. Thus, $\Delta\dot\varpi$ can be used to put constraints on the parameters like ${\ell}$ entering the exotic models one is interested in. From \rfr{precess} it turns out that the tightest constraints come from Mercury, which is the innermost planet with $a=0.38$ AU, for $k=1$.

 Fienga et al. \cite{fienga}, who used also a few data from the three flybys of MESSENGER in 2008-2009, released an uncertainty of $0.6$ mas cty$^{-1}$ for the perihelion precession of Mercury, so that it is ${\ell}\lesssim 34-48\ {\rm km}$ for $k=1-1/2$. The uncertainty in the pre-MESSENGER Mercury's perihelion extra-precession by Pitjeva \cite{pitjeva} is about one order of magnitude larger (5 mas cty$^{-1}$).

Slightly tighter constraints on ${\ell}$ come from the interplanetary Earth-Mercury ranging. Indeed, according to Table 1 of Ref. \cite{fienga}, the standard deviation $\sigma_{\Delta\rho}$ of the Mercury range residuals $\Delta\rho$, obtained with the INPOP10a ephemerides and including also three Mercury MESSENGER flybys in 2008-2009, is as large as $1.9$ m. A numerical integration of the equations of motion of Mercury and the Earth yields a RS range signal with the same standard deviation for ${\ell}\lesssim 13-18\ {\rm km}\ (k=1-1/2)$. Figure \ref{figura2} displays the case $k=1/2$.
\begin{figure*}
\epsfig{file=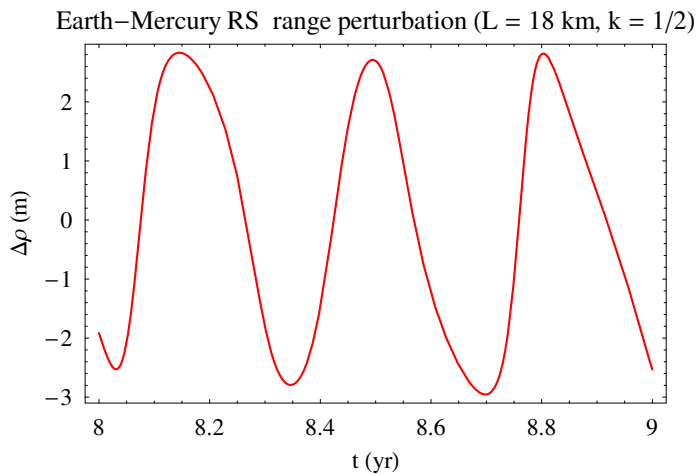}
\caption{2008-2009 Earth-Mercury range shift $\Delta\rho$, in m, due to \textcolor{black}{an unmodeled  potential like} \rfr{upert} for ${\ell}=18$ km and $k=1/2$ computed from the difference between  two numerical integrations of the equations of motion in cartesian coordinates  performed with and without \rfr{upert}. Both the integrations share the same initial conditions retrieved from the NASA WEB interface HORIZONS at http://ssd.jpl.nasa.gov/horizons.cgi, for the epoch J2000.0. It is $\left\langle \Delta\rho \right\rangle=-0.04\ {\rm m},\ \sigma_{\Delta\rho}=1.9\ {\rm m}$. }\lb{figura2}
\end{figure*}
It turns out that the residuals of right ascension (RA) and declination (DEC) in Ref. \cite{fienga} yield much less tight constraints.

As far as natural bodies of the solar system are concerned, the Moon yields better results. Its orbit is accurately reconstructed with the Lunar Laser Ranging (LLR) technique \cite{LLR} since 1969; Figure B-1 of Ref. \cite{folkner} shows that the residuals
of the Earth-Moon range are at a cm-level since about 1990. The secular precession of the lunar perigee is known with an accuracy of about $0.1\ {\rm mas\ yr}^{-1}$ \cite{accura1,accura2,accura3,accura4}, so that \rfr{precess} yields ${\ell}\lesssim 524-741\ {\rm m}\ (k=1-1/2)$.
\subsection{Constraints from the GRACE spacecraft orbiting the Earth}\lb{grazia}
Remaining within the solar system, tighter constraints can be obtained from selected spacecrafts orbiting the Earth.
The Gravity Recovery and Climate Experiment (GRACE) mission \cite{grace}, jointly launched in March 2002 by NASA and the German Space Agency (DLR) to  map the terrestrial gravitational field with an unprecedented accuracy, consists of a tandem of two spacecrafts moving along low-altitude, nearly polar orbits  continuously linked by a Satellite to Satellite Tracking (SST) microwave K-band ranging (KBR) system accurate to better than $10\ \mu$m (biased range $\rho$) \cite{range} and $1\ \mu$m s$^{-1}$ (range-rate $\dot\rho$) \cite{range,rrate}. Studies for a follow-on of GRACE show that the use of a  interfermoteric laser ranging system may push the accuracy in the range-rate to a $\sim 0.6\ {\rm nm\ s^{-1}}$ level. A numerical integration of the equations of motion for GRACE A/B, including also the mismodelled signal of the first nine zonal harmonics of geopotential \cite{capderou} according to the global Earth gravity model GOCO01S \cite{goco01s}, shows that the SST range is more effective than the SST range-rate in constraining the RS parameter for which it holds ${\ell}\lesssim 5\ {\rm m}$. Figure \ref{figura3} depicts the numerically integrated GRACE SST range signal due to \rfr{upert} and the aforementioned mismodeled zonals.
\begin{figure*}
\epsfig{file=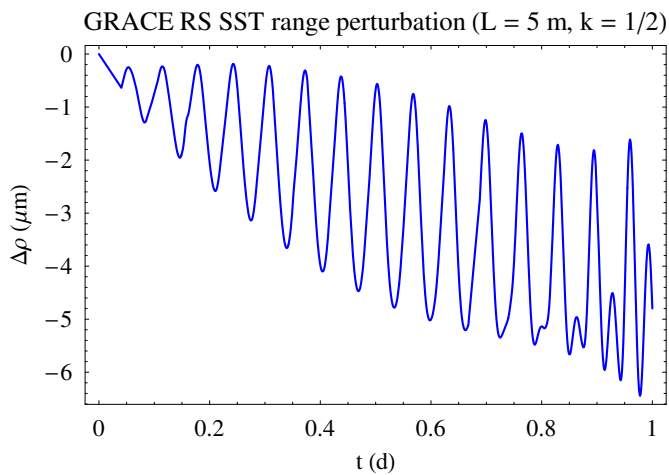}
\caption{Daily GRACE SST range shift $\Delta\rho$, in $\mu$m, due to \textcolor{black}{an unmodeled  potential like} \rfr{upert} (${\ell}=5$ m, $k=1/2$) and the first nine mismodelled zonals of geopotential according to GOCO01S \cite{goco01s} computed from the difference between  two numerical integrations of the equations of motion in cartesian coordinates  performed with and without the perturbations considered. Both the integrations share the same initial conditions contained in the files GNV1B$\_$2003-09-14$\_$A$\_$00 and  GNV1B$\_$2003-09-14$\_$B$\_$00 retrieved from ftp://cddis.gsfc.nasa.gov/pub/slr/predicts/current/graceA$\_$irvs$\_$081202$\_$0.gfz and ftp://cddis.gsfc.nasa.gov/pub/slr/predicts/current/graceB$\_$irvs$\_$081201$\_$1.gfz. The epoch is 13 September 2003. (See ftp://podaac.jpl.nasa.gov/pub/grace/doc/Handbook$\_$1B$\_$v1.3.pdf for the explanation of the GPS Navigation Data Format Record (GNV1B) format).
It is $\left\langle \Delta\rho \right\rangle=-3\ \mu{\rm m},\ \sigma_{\Delta\rho}=2\ \mu{\rm m}$. }\lb{figura3}
\end{figure*}
It can be shown that a sub-nm s$^{-1}$ level of accuracy in the GRACE SST range-rate would imply the possibility of constraining ${\ell}$ down to $\sim 10$ cm level.
\subsection{Constraints from the double pulsar and Sgr A$^{\ast}$}\lb{neutrona}
Constraints comparable with the planetary ones can be obtained from the periastron of the double pulsar PSR J0737-3039A/B system \cite{burgay,lyne}.
Indeed, the semimajor axis of its relative orbit amounts to just $a=0.006$ AU \cite{kramer}. The present-day accuracy in measuring the secular precession of the periastron is $6.8\times 10^{-4}$ degree per year (deg yr$^{-1}$ in the following) \cite{kramer}; thus, a straightforward application of it to \rfr{precess} would give ${\ell}\lesssim 11-16\ {\rm km}\ (k=1-1/2)$. Actually, the larger uncertainty in the theoretical expression of the general relativistic 1PN periastron precession must be taken into account as well. It is as large as $0.03$ deg yr$^{-1}$ \cite{iorio}, so that it yields ${\ell}\lesssim 80-112\ {\rm km}\ (k=1-1/2)$.

The perinigricon of the S2 star, orbiting in $15.98$ yr the compact object hosted in Sgr A$^{\ast}$ with $M=4\times 10^6 M_{\odot}$ \cite{ghez} at $\left\langle r\right\rangle = 1433$ AU from it \cite{gillessen}, yields much weaker constraints on the asymptotic AdS curvature. Indeed, an accuracy of \cite{gillessen}
$0.05\ {\rm deg\ yr}^{-1}$ can be inferred for the perinigricon precession of S2: thus, ${\ell}\lesssim 6.2-8.8\ {\rm AU}$.
\textcolor{black}{\section{Model-independent constraints on $r^{-3}$ interactions}
The validity of the previous results is not necessarily limited just to the RS braneworld model. Indeed, they are, in fact, quite model-independent in the sense that they hold for any long-range modification $U_{\rm pert}\propto r^{-3}$ \cite{moste,ferrer,ferrer2,dob,adelberger,adel09} of the usual Newtonian potential. In particular, $\ell$ must not necessarily be identified with the AdS radius of curvature of the RS braneworld model.}

\textcolor{black}{By assuming that the putative, new interaction does not depend on the specific matter distribution generating the gravitational field, a  universal parameter can be introduced with the replacement
\eqi kGM{\ell^2}\rightarrow \mathcal{K},\ [\mathcal{K}]={\rm L}^5\ {\rm T}^{-2}.\lb{zolf}\eqf
Thus, the bounds on $\ell$ of Section \ref{pianeti}-Section \ref{neutrona} straightforwardly translate into constraints on $\mathcal{K}$ itself.}

\textcolor{black}{It turns out that the tightest bounds come from the Earth-GRACE system yielding $\mathcal{K}\leq 1\times 10^{16}\ {\rm m^5\ s^{-2}}$.}
\section{Summary and conclusions}
Table \ref{tavola} \textcolor{black}{ and Table \ref{tavola2} } resume our findings.
\begin{table*}
\caption{\label{tavola}Summary of the upper bounds on $\ell$ inferred from the various scenarios discussed in the paper. In each space of the row for $\ell_{\rm max}$, the first number from the left refers to $k=1$, while the second one refers to $k=1/2$.
}
\begin{ruledtabular}
\begin{tabular}{lllllll}
 & Mercury ($\Delta\dot\varpi$) & Mercury ($\Delta\rho$) & Moon ($\Delta\dot\varpi$) & GRACE ($\Delta\rho$) & PSR J0737-3039A/B ($\Delta\dot\varpi$) & S2-Sgr A$^{\ast}$ ($\Delta\dot\varpi$)\\
\colrule
$\ell_{\rm max}$ & $34-48$ km & $13-18$ km & $524-741$ m  & $5$ m  & $80-112$ km  & $6.2-8.8$ AU\\
\end{tabular}
\end{ruledtabular}
\end{table*}
\begin{table*}
\textcolor{black}{
\caption{\label{tavola2}Summary of the upper bounds, in ${\rm m^5\ s^{-2}}$, on the model-independent parameter $\mathcal{K}$ of a $\propto r^{-3}$ modification of the $r^{-1}$ Newtonian potential inferred from the various scenarios discussed in the paper. They have been obtained from the values of $\ell_{\rm max}$ in Table \ref{tavola}  with the replacement $k GM {\ell^2}\rightarrow {\mathcal{K}}$. $\mathcal{K}$ has been assumed universal.
}
\begin{ruledtabular}
\begin{tabular}{lllllll}
 & Mercury ($\Delta\dot\varpi$) & Mercury ($\Delta\rho$) & Moon ($\Delta\dot\varpi$) & GRACE ($\Delta\rho$) & PSR J0737-3039A/B ($\Delta\dot\varpi$) & S2-Sgr A$^{\ast}$ ($\Delta\dot\varpi$)\\
\colrule
$\mathcal{K}_{\rm max}$ & $1.5\times 10^{29}$  & $2\times 10^{28}$  & $1\times 10^{20}$   & $1\times 10^{16}$   & $1\times 10^{30}$   & $5\times 10^{50}$ \\
\end{tabular}
\end{ruledtabular}
}
\end{table*}

\textcolor{black}{As far as the $\ell^2$ form of the $r^{-3}$ extra-potential is concerned, Table \ref{tavola} tells us that} tightest bounds, at the m level, come from the Satellite-to-Satellite Tracking ranging between the GRACE spacecrafts.

\textcolor{black}{If we interpret them in terms of the RS model,} from a general point of view, we find \textcolor{black}{not entirely adequate} arguing that Earth-based laboratory-scale tests of several modified models of gravity are in all respects superior since they can deliver much tighter constraints. Indeed, apart from the basic fact that it is important to scrutiny a theoretical paradigm in different, complementary scenarios, some authors also pointed out that there are theoretical reasons for wanting to test such foundational issues of gravity in space since extra gravitational degrees of freedom could have an environmental dependence.
Moreover, about the seemingly superiority of certain constraints of the RS model inferred from some astrophysical systems,  often they rely upon more or less speculative and untested theoretical assumptions,  and  the phenomena adopted
may depend on the system's composition, formation and dynamical history as well. Thus, we reiterate the importance of finding solid, well-understood means of constraining $\ell$ like those proposed here.
We also point out that the same method introduced here could yield
another factor of $10^2$ improvement, constraining $\ell < 10$ cm, provided that the
proposed improvement to satellite-to-satellite tracking is implemented
for the follow-on of the GRACE mission.

\textcolor{black}{Last but not least}, we remark that, actually, our analysis is not necessarily limited to the RS model since it is valid for whatsoever theoretical scheme predicting $r^{-3}$ corrections to the Newtonian potential. \textcolor{black}{In this respect, the occurrence of a characteristic length scale $\ell$ is common to a wide class of power-law interactions, so that it should not necessarily be thought of as the AdS radius of curvature in the RS model. Thus, the bounds of Table \ref{tavola} have a wider range of applicability. Table \ref{tavola2} shows that GRACE yields the tightest bounds on a universal coupling parameter $\mathcal{K}$ which is constrained  to  $\leq 10^{16}\ {\rm m^5\ s^{-2}}$. It could be pushed down to a $10^{12} \ {\rm m^5\ s^{-2}}$ level by  GRACE follow-on.}
%
%


\end{document}